\documentclass[12pt]{article}
\usepackage[centertags]{amsmath}
\usepackage{amssymb,amsfonts}
\usepackage[comma,super]{natbib}
\usepackage{latexsym}
\newcommand{\be}{\begin{equation}} \newcommand{\ee}{\end{equation}}
\newcommand{\bea}{\begin{eqnarray}} \newcommand{\eea}{\end{eqnarray}}
\newcommand{\bse}{\begin{subequations}}\newcommand{\ese}{\end{subequations}}
\newcommand{\n}{\nonumber}
\begin{document}

\title{\textbf{Analytical models for quark stars}}
\author{K. Komathiraj\thanks{Permanent
address: Department of Mathematical Sciences, South Eastern
University, Sammanthurai, Sri Lanka.}\; and S. D.
Maharaj\thanks{eMail:
\texttt{maharaj@ukzn.ac.za}}\\
Astrophysics and Cosmology Research Unit,\\ School of Mathematical
Sciences, University of KwaZulu-Natal,\\Private Bag X54001,
 Durban 4000, South Africa.}
\date{}
\maketitle
\begin{abstract}
We find two new classes of exact solutions to the Einstein-Maxwell
system of equations. The matter content satisfies a linear
equation of state consistent with quark matter; a particular form
of one of the gravitational potentials is specified to generate
solutions. The exact solutions can be written in terms of
elementary functions, and these can be related to quark matter in
the presence of an electromagnetic field. The first class of
solutions generalises the Mak and Harko model. The second class of
solutions does not admit any singularities in the matter and
gravitational potentials at the centre.

 \vspace{0.5cm} \noindent
\end{abstract}

\section{Introduction}
The existence of quark stars in hydrostatic equilibrium was first
suggested by Itoh\cite{Itoh} in a seminal treatment. The study of
strange stars consisting of quark matter has stimulated much
interest in the last few decades since this could represent the most
energetically favourable state of baryon matter. Matter consisting
of u, d and s quarks may be the absolute ground state of matter at
zero pressure and temperature as suggested by Bodmer\cite{Bod}. It
is expected that strange stars form during the collapse of the core
of a massive star after a supernova explosion\cite{Che}. In regions
of low temperatures and sufficiently high densities hadrons are
crushed into quark matter with color superconducting phases  which
occur in the dense cores of neutron stars as remarked by
Alford\cite{Alf}. Consequently the core of a proto-neutron star or
neutron star provides the appropriate environment for  ordinary
matter to convert to strange quark matter. Another possibility is
that a rapidly spinning dense star can accrete sufficient mass to
undergo a phase transition to form a strange star.

As the physics of ultrahigh densities for quark matter is not well
understood, researchers restrict their attention to the
phenomenological MIT bag model\cite{Cho}$^{-}$\cite{Wit}. In the
bag model, the strange matter equation of state has a simple
linear form given by \be\label{eq:1} p=\frac{1}{3}(\rho-4B)\ee
where $\rho$ is the energy density, $p$ is the isotropic pressure
and $B$ is the bag constant. The quark confinement is determined
by the vacuum pressure $B$ in the bag model that equilibrates the
pressure of quarks thereby stabilising the system. Studies of
particular compact astronomical objects indicate that they could
be strange stars with the quark matter\cite{Bom}$^{-}$\cite{Uso}.
A candidate for a strange star may have been observed using the
deep Chandra LETG+HRC-S observations; Drake \emph{et al}\cite{Dra}
suggested that the X-ray source RXJ1856.5-3754 may be such an
object. Sotani \emph{et al}\cite{Sot} have used observational data
on gravitational waves to obtain the equation of state for quark
matter. Harko and Cheng\cite{Har} considered collapsing strange
matter in spherically symmetric fields. Yilmaz and
Baysal\cite{Yil} studied charged strange matter in rotating
fields. The role of anisotropy, with the linear equation of state
(\ref{eq:1}), was pursed by Mak and Harko\cite{Mak1} and Sharma
and Maharaj\cite{Sha} who demonstrated exact analytical solutions.

In a recent treatment Mak and Harko\cite{Mak2} found a charged
strange quark star under the assumption of spherical symmetry and
the existence of a conformal Killing vector. In this paper we
consider the Einstein-Maxwell system of equations with the linear
equation of state (\ref{eq:1}) and apply them to strange stars.
The existence of a conformal symmetry is not an assumption that we
make. We demonstrate that exact analytical solutions to the field
equations are possible that contain the Mak-Harko
model\cite{Mak2}. In Section 2, we rewrite the Einstein-Maxwell
field equations for the static spherically line element as an
equivalent set of differential equations utilising a
transformation due to Durgapal and Bannerji\cite{Dur}. We then
obtain a new set of differential equations with the assistance of
the bag equation of state (\ref{eq:1}) for strange matter. On
specifying an explicit form for one of the gravitational
potentials, we obtain a first order differential equation in the
remaining potential in Section 3. In Section 4 we find a new class
of exact solutions to the Einstein-Maxwell system. The model of
Mak and Harko \cite{Mak2} is regained as a special case. In
Section 5 we present a second class of exact solutions that
satisfy the Einstein-Maxwell system. This category of solutions
has the desirable feature of not admitting singularities at the
centre.

\section{Spherically symmetric spacetimes}
Since our intention is to study relativistic stellar objects it
seems reasonable, on physical grounds, to assume that spacetime is
static and spherically symmetric. This is clearly consistent with
models utilised to study physical processes in compact objects.
The metric of a static spherically symmetric spacetime in
curvature coordinates reads \be \label{eq:2}
ds^{2}=-e^{2\nu(r)}dt^{2}+e^{2\lambda(r)}dr^{2}+r^{2}(d\theta^{2}+\sin^{2}\theta
d\phi^{2})\ee where $\nu(r)$  and $\lambda(r)$ are the two
arbitrary  functions. For charged perfect fluids the
Einstein-Maxwell system of field equations are obtained
as\bse\label{eq:3}\bea
\frac{1}{r^{2}}(1-e^{-2\lambda})+\frac{2\lambda^\prime}{r}e^{-2\lambda}&=&\rho+\frac{1}{2}E^{2}\\
-\frac{1}{r^{2}}(1-e^{-2\lambda})+\frac{2\nu^\prime}{r}e^{-2\lambda}&=&p-\frac{1}{2}E^{2}\\
e^{-2\lambda}\left(\nu^{\prime\prime}+{\nu^\prime}^2+\frac{\nu^\prime}{r}-\nu^\prime\lambda^\prime-\frac{\lambda^\prime}{r}\right)&=&p+\frac{1}{2}E^{2}\\
\sigma&=&\frac{1}{r^{2}}e^{-\lambda}(r^{2}E)^\prime\eea\ese for the
line element (\ref{eq:2}). We are utilising units in which the
coupling constant $8\pi G/c^4 =1$ and the speed of light $c=1$. The
energy density $\rho$ and the pressure $p$ are measured relative to
the comoving fluid 4-velocity $u^{a}=e^{-\nu}\delta^{a}_{0}$ and
primes means a derivative with respect to the radial coordinate $r$.
In the system (\ref{eq:3}) the quantities $E$ and $\sigma$ are the
electric field intensity and the proper charge density respectively.

It is convenient at this point to introduce the transformation
\be\label{eq:4}
A^{2}y^{2}(x)=e^{2\nu(r)},~~~Z(x)=e^{-2\lambda(r)},~~~x=Cr^{2}\ee
where $A$ and $C$ are arbitrary constants. With this
transformation, the system (\ref{eq:3}) has the equivalent form
 \bse\label{eq:5} \bea
\label{eq:5a}\frac{1-Z}{x}-2\dot{Z}&=&\frac{\rho}{C}+\frac{E^{2}}{2C}\\\n\\
\label{eq:5b}4Z\frac{\dot{y}}{y}+\frac{Z-1}{x}&=&\frac{p}{C}-\frac{E^{2}}{2C}\\\n\\
\label{eq:5c}4Zx^{2}\ddot{y}+2\dot{Z}x^{2}\dot{y}+\left(\dot{Z}x-Z+1-\frac{E^{2}x}{C}\right)y&=&0\\\n\\
\label{eq:5d}\frac{\sigma^{2}}{C}&=&\frac{4Z}{x}(x\dot{E}+E)^{2}\eea\ese
where dots denote differentiation with respect to the variable
$x$. Note that equation (\ref{eq:5c}) is the condition of pressure
isotropy. We can replace the system of field equations
(\ref{eq:5}), including the bag equation of state (\ref{eq:1}), by
the system
\bse\label{eq:6}\bea\label{eq:6a} \rho&=&3p+4B\\
\label{eq:6b}\frac{p}{C}&=&Z\frac{\dot{y}}{y}-\frac{1}{2}\dot{Z}-\frac{B}{C}\\
\label{eq:6c}\frac{E^{2}}{2C}&=&\frac{1-Z}{x}-3Z\frac{\dot{y}}{y}-\frac{1}{2}\dot{Z}-\frac{B}{C}\\
\label{eq:6d}0&=&4Zx^{2}\ddot{y}+(6xZ+2x^{2}\dot{Z})\dot{y}+\left[2x\left(\dot{Z}+\frac{B}{C}\right)+Z-1\right]y\\
\sigma&=&2\sqrt{\frac{CZ}{x}}(E+x\dot{E})\eea\ese The system of
equations (\ref{eq:6}) governs the gravitational behaviour of a
charged quark star.
\section{Integration procedure}
We describe one possible integration procedure that leads to an
exact solution of the Einstein-Maxwell system (\ref{eq:6}). Note
that other procedures are possible; our approach has the advantage
of producing a first order equation that has solution in terms of
elementary functions. We observe from (\ref{eq:6a}) that $\rho$ and
$p$ are related. Therefore in the system (\ref{eq:6}) there are five
independent variables $(Z,~y,~p~\textrm{or}~ \rho,~E,~\sigma )$ and
only four independent equations. We have freedom to choose only one
of the quantities involved. In our approach we specify $y(x)$ on
physical grounds. A number of choices for the gravitational
potential $y(x)$ are possible; clearly we should choose a form that
is likely to lead to a physically reasonable solution. To make the
above set of equations tractable, we choose the metric function in
the particular form \be \label{eq:7} y(x)=(a+x^{m})^{n}\ee where
$a,m$ and $n$ are constants. The form chosen ensures that the metric
function $y$ is continuous and well-behaved in the interior of the
star for the wide range of values of parameters $m$ and $n$. The
function $y$
 yields a finite value at the centre of the star. This is a very desirable
 feature for the model on physical grounds. It is
interesting to observe that many of the solutions found previously
do not share this feature.

Substitution of (\ref{eq:7}) into (\ref{eq:6d}) leads to the first
order equation \bea
&&\dot{Z}+\frac{a^{2}+2a[1+mn(2m+1)]x^{m}+[2mn(2mn+1)+1]x^{2m}}{2x(a+x^{m})[a+(1+mn)x^{m}]}Z\n\\
&=&\frac{(1-\frac{2B}{C}x)(a+x^{m})}{2x[a+(1+mn)x^{m}]} \n\eea
This first order equation is linear so that it can be integrated
in principle. The complicated rational coefficient of $Z$ can be
simplified using partial fractions and we obtain \bea\label{eq:8}
&&\dot{Z}+\left[\frac{1}{2x}+\frac{2m(n-1)x^{m-1}}{(a+x^{m})}+\frac{m[4(1+mn)-3n]x^{m-1}}{2[a+(1+mn)x^{m}]}\right]Z\n\\
&=&\frac{(1-\frac{2B}{C}x)(a+x^{m})}{2x[a+(1+mn)x^{m}]}\eea Note
that we have essentially reduced the solution of the field
equations (\ref{eq:6}) to integrating (\ref{eq:8}). Once the
potential $Z$ in (\ref{eq:8}) is found the remaining relevant
quantities $\rho,~p$ and $E$ then follow from (\ref{eq:6a}),
(\ref{eq:6b}) and (\ref{eq:6c}) respectively. It is possible to
find exact solutions to the Einstein-Maxwell field equations with
the linear equation of state for different
 values of $m$ and $n$ in (\ref{eq:8}). We illustrate this with
 two simple examples in terms of elementary functions. Other exact
 solutions are possible but the form of the solution becomes more
 complicated and could involve special functions.
\section{Generalised Mak-Harko model}
An exact solution of (\ref{eq:6}) can be found with $m=1/2$ and
$n=1$. In this case (\ref{eq:7}) gives the first metric function
\be y(x)=(a+\sqrt{x})\n\ee Equation (\ref{eq:8}) becomes
\be\label{eq:9}
\dot{Z}+\left[\frac{1}{2x}+\frac{3}{2\sqrt{x}(2a+3\sqrt{x})}\right]Z=\frac{(1-\frac{2B}{C}x)(a+\sqrt{x})}{x(2a+3\sqrt{x})}\n\ee
which can be integrated to give the second metric function
 \be\label{eq:10}
Z=\frac{3(2a+\sqrt{x})-\frac{B}{C}x(4a+3\sqrt{x})}{3(2a+3\sqrt{x})}\ee
Hence we can generate the exact analytical model
 \bse \label{eq:11}\bea
e^{2\nu}&=&A^{2}(a+\sqrt{x})^{2}\\
e^{2\lambda}&=&\frac{3(2a+3\sqrt{x})}{3(2a+\sqrt{x})-\frac{B}{C}x(4a+3\sqrt{x})}\\
\rho&=&f(x)+\frac{B(16a^{3}+47a^{2}\sqrt{x}+48ax+18x^{\frac{3}{2}})}{2(a+\sqrt{x})(2a+3\sqrt{x})^{2}}\\
3p&=&f(x)-\frac{B(16a^{3}+81a^{2}\sqrt{x}+120ax+54x^{\frac{3}{2}})}{2(a+\sqrt{x})(2a+3\sqrt{x})^{2}}\\
E^{2}&=&\frac{C(-2a^{2}-2a\sqrt{x}+3x)+Bx(a^{2}+2a\sqrt{x})}{\sqrt{x}(a+\sqrt{x})(2a+3\sqrt{x})^{2}}\eea\ese
which is a solution of (\ref{eq:6}). For simplicity we have set
 \be
f(x)=\frac{3C(6a^{2}+10a\sqrt{x}+3x)}{2\sqrt{x}(a+\sqrt{x})(2a+3\sqrt{x})^{2}}\n\ee

The exact model (\ref{eq:11}) satisfies the Einstein-Maxwell
system (\ref{eq:6}). Note that if we set $a=0$, then the system
(\ref{eq:11}) becomes \be \label{eq:12}
e^{2\nu}=D^{2}r^{2},~e^{2\lambda}=\frac{3}{1-Br^{2}},~\rho=\frac{1}{2r^{2}}+B,~p=\frac{1}{6r^{2}}-B,~E^{2}=\frac{1}{3r^{2}}\ee
where we have set $D^{2}=A^{2}C$. The particular solution
(\ref{eq:12}) was found by Mak and Harko\cite{Mak2} under the
assumption of spherical symmetry and the existence of a
one-parameter group of conformal motions. It is interesting to
observe that on substituting the values $a=0$ and $B=0$ for the
constants, the solution (\ref{eq:11}) becomes identical to that
obtained by Misner and Zapolsky \cite{Mis}. The physical features
of the solutions (\ref{eq:12}) were studied by Mak and Harko
\cite{Mak2} and shown to be consistent with the interior of a
quark star with charged material. This corresponds to a single
stable quark configuration with radius $R=9.46~ Km$ and mass
$M=2.86 M_{\odot}$; these figures are consistent with values
obtained using numerical methods by other researchers
\cite{Hae1}$^{-}$\cite{Gou}. Consequently our more general class
of solutions is likely to produce charged quark models consistent
with stellar evolution and observational data. We comment that our
new class of solutions (\ref{eq:11}) has a singularity in the
charge density and mass density at the centre; the Mak and Harko
\cite{Mak2} model also shares this feature. The singularity in the
charge density and mass density is physically acceptable since the
total charge and mass remain finite. However our gravitational
potentials $e^{2\nu}$ and $e^{2\lambda}$ remain finite at the
centre which contrasts with the singularities in the metric
functions of Mak and Harko when $x=0$.

\section{Nonsingular quark model}
Another exact solution of (\ref{eq:6}) can be found with $m=1$ and
$n=2.$ For this case (\ref{eq:7}) gives the first metric function
\be y(x)=(a+x)^{2}.\n\ee Equation (\ref{eq:8}) becomes
\be\label{eq:13}
\dot{Z}+\left[\frac{1}{2x}+\frac{2}{a+x}+\frac{3}{a+3x}\right]Z=\frac{(1-\frac{2B}{C}x)(a+x)}{2x(a+3x)}\n\ee
which can be integrated to give the second metric function
\be\label{eq:14}Z=\frac{9(35a^{3}+35a^{2}x+21ax^{2}+5x^{3})-\frac{2B}{C}x(105a^{3}+189a^{2}x+135ax^{2}+35x^{3})}{315(a+x)^{2}(a+3x)}\ee
Therefore we can find the exact analytical model
\bse\label{eq:15}\bea
e^{2\nu}&=&A^{2}(a+x)^{4}\\\n\\
e^{2\lambda}&=&\frac{315(a+x)^{2}(a+3x)}{9(35a^{3}+35a^{2}x+21ax^{2}+5x^{3})-\frac{2B}{C}x(105a^{3}+189a^{2}x+135ax^{2}+35x^{3})}\n\\\\\n\\
\rho&=&
g(x)+\frac{2B[3(35a^{5}+133a^{4}x+246a^{3}x^{2})+5(254a^{2}x^{3}+209ax^{4}+63x^{5})]}{105(a+x)^{3}(a+3x)^{2}}\\\n\\\n\\
3p&=&g(x)-\frac{2B[3(35a^{5}+497a^{4}x+1854a^{3}x^{2})+5(1678a^{2}x^{3}+1177ax^{4}+315x^{5}]}{105(a+x)^{3}(a+3x)^{2}}\n\\
&=&(\rho-4B)\\\n\\
E^{2}&=&4x[9C(49a^{3}+363a^{2}x+339ax^{2}+105x^{3})\n\\
&-&2B(21a^{4}+162a^{3}x+816a^{2}x^{2}+910ax^{3}+315x^{4})]/315(a+x)^{3}(a+3x)^{2}
\eea\ese  Again for simplicity we have set  \be
g(x)=\frac{6C(70a^{4}+217a^{3}x+159a^{2}x^{2}+75ax^{3}+15x^{4})}{35(a+x)^{3}(a+3x)^{2}}\n\ee

The exact model (\ref{eq:15}) satisfies the Einstein-Maxwell
system (\ref{eq:6}) and constitutes a new family of analytical
solutions for a quark star with charged material. The
gravitational potentials $e^{2\nu}$ and $e^{2\lambda}$ in
(\ref{eq:15}) have the advantage of having a simple analytic form,
and they are written in terms of polynomials and rational
functions. Consequently  the matter variables and the electric
field intensity have a simple analytic representation. The
function $e^{2\nu}$ is continuous and well behaved in the interior
and finite at the centre $x=0.$ The function $e^{2\lambda}$ is
well behaved and has a constant value  at the centre $x=0$. The
energy density $\rho$ is positive throughout the interior, regular
at the centre with value $\rho_{0}=2(\frac{6C}{a}+B)$. The
pressure $p$ is regular at the centre with value
$p_{0}=2(\frac{2C}{a}-\frac{B}{3})=\frac{1}{3}(\rho_{0}-4B)$. The
electric field intensity $E$ is continuous in the interior and
vanishes at the centre. Hence the matter variables and
gravitational potentials comply with usual conditions for a
stellar source. The finiteness of
$e^{2\nu},~e^{2\lambda},~\rho,~p$ and $E$ at the origin $x=0$ is a
very welcome feature which is absent in the previous class of
solutions. Consequently the exact solutions (\ref{eq:15}) are
likely to produce charged quark stars with physically acceptable
interiors. A recent attempt in this direction is the strange star
model of Jotania and Tikekar \cite{Jot} admitting compact
configurations with mass to size ratio consistent with strange
matter.

\section{Conclusion}
We have generated a new category of exact solutions to the
Einstein-Maxwell system of equations. The linear equation of state
(\ref{eq:1})  was imposed which is relevant in the description of
quark stars. Two classes of exact solutions were identified by
specifying the form of one of the gravitational potentials. The
first class comprises the generalised Mak-Harko model as the earlier
solution of Mak and Harko\cite{Mak2} is regained as a special case.
The second class comprises the nonsingular quark model which has
finite values for both the matter and metric variables at the
stellar centre. The method in this paper depends crucially on the
choice (\ref{eq:7}) for the metric potential $Z$ which leads to
solutions of the condition of pressure isotropy. In future work it
would be desirable to seek physically reasonable solutions, with new
forms of $Z$, which are consistent with the relationship
(\ref{eq:1}). We also intend to study more closely the physical
features of quark stars, from the exact solutions generated, and
relate these to specific astronomical objects following the
treatment of Jotania and Tikekar\cite{Jot}.

In conclusion, we make two further points that are of significance
in the modelling process. Firstly, in the general solutions
(\ref{eq:11}) and (\ref{eq:15}), when studying models of charged
spheres, we should consider only those values of parameters for
which the energy density $\rho$, the pressure $p$ and the electric
field intensity $E^2$ are positive. Clearly a wide range of charged
spheres, with nonsingular potentials and matter variables, are
possible for relevant choices of $a, B$ and $C$. Secondly, the
interior metric (\ref{eq:2}) must match to the Reissner-Nordstrom
exterior spacetime
\[
 d s^2=-\left( 1-\frac{2M}{r} + \frac{Q^2}{r^2}\right) d t^2 +
 \left( 1-\frac{2M}{r} + \frac{Q^2}{r^2}\right)^{-1} d r^2 +
r^2( d \theta^2 + \sin^2 \theta d \phi^2)
\]
across the boundary of the star. This generate relationships between
the mass $M$, the charge $Q$, and the constants $a,B$ and $C$.
Consequently the continuity of the metric coefficients across the
boundary of the star is easily achieved.  There are sufficient free
parameters available to satisfy the necessary conditions that may
arise from a particular physical model under consideration.
\\

\noindent{\large \bf Acknowledgements}\\

\noindent KK thanks the National Research Foundation and the
University of KwaZulu-Natal for financial support, and also
extends his appreciation to the South Eastern University of Sri
Lanka for granting study leave. SDM acknowledges that this work is
based upon research supported by the South African Research Chair
Initiative of the Department of Science and Technology and the
National Research Foundation.

%
%---------------end of bibliography----------------------------
%
%


\begin{thebibliography}{0}
%------------------------------------------------------------

\bibitem{Itoh} N. Itoh, \emph{Prog. Theor. Phys.} \textbf{44},
291 (1970).

\bibitem{Bod}
A. R. Bodmer, \emph{Phys. Rev. D} \textbf{4}, 1601 (1971).

\bibitem{Che}
K. S. Cheng, Z. G. Dai and T. Lu, \emph{Int. J. Mod. Phys. D}
\textbf{7}, 139 (1998).

\bibitem{Alf}
M. Alford, \emph{Ann. Rev. Nucl. Part. Sci.} \textbf{51}, 131
(2001).

\bibitem{Cho}
A. Chodos, R. L. Jaffe, K. Johnson, C. B. Thorn and V. F.
Weisskopf, \emph{Phys. Rev. D} \textbf{9}, 3471 (1974).

\bibitem{Far}
E. Farhi and R. L. Jaffe, \emph{Phys. Rev. D} \textbf{30}, 2379
(1984).

\bibitem{Wit}
E. Witten, \emph{Phys. Rev. D} \textbf{30}, 272 (1984).

\bibitem{Bom}
I. Bombaci, \emph{Phys. Rev. C} \textbf{55}, 1587 (1997).

\bibitem{Li1}
X.-D Li, Z.-G Dai and Z.-R Wang, \emph{Astron. Astrophys.}
\textbf{303}, L1 (1995).

\bibitem{Dey}
M. Dey, I. Bombaci, J. Dey, S. Ray and B. C. Samanta, \emph{Phys.
Lett. B} \textbf{438}, 123 (1998).

\bibitem{Li2}
X.-D. Li, I. Bombaci, M. Dey, J. Dey and E. P. J. van den Heeuvel,
\emph{Phys. Rev. Lett.} \textbf{83}, 3776 (1999).

\bibitem{Li3}
X.-D. Li, S. Ray, J. Dey, M. Dey and I. Bombaci, \emph{Astrophys.
J.} \textbf{527}, L51 (1999).

\bibitem{Xu1}
R. X. Xu, G. J. Qiao and B. Zhang, \emph{Astrophys. J.}
\textbf{522}, L109 (1999).

\bibitem{Xu2}
R. X. Xu, X. B. Xu and X. J. Wu, \emph{Chin. Phys. Lett.}
\textbf{18}, 837 (2001).

\bibitem{Pon}
J. A. Pons, F. M. Walter, J. M. Lattimer, M. Prakash, R. Neuhauser
and A. Penghui, \emph{Astrophys. J.} \textbf{564}, 981 (2002).

\bibitem{Uso}
V. V. Usov, \emph{Phys. Rev. D} \textbf{70}, 067301 (2004).

\bibitem{Dra}
J. J. Drake, H. L. Marshall, S. Dreizler, P. E. Freeman, A.
Fruscione, M. Juda, V. Kashyap, F. Nicastro, D. O. Pease, B. J.
Wargelin and K. Werner, \emph{Astrophys. J.} \textbf{572}, 996
(2002).

\bibitem{Sot}
H. Sotani, K. Kohri and T. Harada, \emph{Phys. Rev. D}
\textbf{69}, 084008 (2004).

\bibitem{Har}
T. Harko and K. S. Cheng, \emph{Phys. Lett. A} \textbf{266}, 249
(2000).

\bibitem{Yil}
I. Yilmaz and H. Baysal, \emph{Int. J. Mod. Phys. D} \textbf{14},
697 (2005).

\bibitem{Mak1}
M. K. Mak and T. Harko, \emph{Chin. J. Astron. Astrophys.}
\textbf{2}, 248 (2002).

\bibitem{Sha}
R. Sharma and S. D. Maharaj, \emph{Mon. Not. R. Astron. Soc.}
\textbf{375}, 1265 (2007).

\bibitem{Mak2}
M. K. Mak and T. Harko, \emph{Int. J. Mod. Phys. D} \textbf{13},
149 (2004).

\bibitem{Dur}
M. C. Durgapal and R. Bannerji, \emph{Phys. Rev. D} \textbf{27},
328 (1983).

\bibitem{Mis}
C. W. Misner and H. S. Zapolsky, \emph{Phys. Rev. Lett.}
\textbf{12}, 635 (1964).

\bibitem{Hae1}
P. Haensel, J. L. Zdunik and R. Schaefer, \emph{Astron.
Astrophys.} \textbf{160}, 121 (1986).

\bibitem{Hae2}
P. Haensel and  J. L. Zdunik, \emph{Nature.} \textbf{340}, 617
(1989).

\bibitem{Gou}
E. Gourgoulhon, P. Haensel, R. Livirne, E. Paluch, S. Bonazzola
and J. A. Marck, \emph{Astron. Astrophys.} \textbf{349}, 851
(1999).

\bibitem{Jot}
K. Jotania and R. Tikekar, \emph{Int. J. Mod. Phys. D}
\textbf{15}, 1175 (2006).
\end{thebibliography}
\end{document}